\documentclass[reprint,aps,prl]{revtex4-1}

\usepackage{graphicx}

\graphicspath{{data/}{figures/}}

\begin{document}
\title{Replica-symmetry breaking for directed polymers}
\author{Alexander K. Hartmann}	
\affiliation{Institut f\"ur Physik, Universit\"at Oldenburg, 26111 Oldenburg,
  Germany}
\begin{abstract}
  Directed polymers on
  1+1 dimensional lattices coupled to a heat bath at temperature $T$
  are studied numerically
  for three ensembles of the 
  site disorder. In particular
  correlations of the disorder as well as fractal patterning 
  are considered. Configurations are directly sampled in
  perfect thermal equilibrium for
  very large system sizes with up to
  $N=L^2= 32768 \times 32768 \approx 10^{9}$ sites.
  The phase-space structure is studied via the distribution of overlaps
  and hierarchical clustering of configurations.
  One ensemble shows a simple behavior like a ferromagnet.
  The other two ensembles exhibit
  indications for  complex behavior reminiscent of
  multiple replica-symmetry
  breaking. Also results for the ultrametricity of the phase space and the
  phase transition behavior of  $P(q)$ when varying the temperature
  $T$ are studied. In total, the present model ensembles
  offer convenient
  numerical accesses to comprehensively  studying complex behavior.  
\end{abstract}

\maketitle

Disordered systems like structural glasses \cite{berthier2011},
spin glasses
\cite{binder1986,mezard1987,fischer1991,young1998,nishimori2001,kawashima2013}
or random optimization
problems \cite{phase-transitions2005,mezard2009,moore2011} exhibit for some
ensembles of disorder realizations complex low-temperature phases,
characterized by rough energy landscapes
and diverging times scales. Most of such models cannot be solved
analytically, except few mean-field ensembles like the
Sherrington-Kirkpatrick (SK) spin-glass \cite{sherrington1975,talagrand2006}.
By solving the SK model,  a particular signature of complex behavior,
\emph{replica-symmetry breaking} (RSB), was introduced
\cite{parisi1979,parisi1983}. Usually, and in the present work,
the term RSB is used
also for other systems exhibiting multi-level hierarchical and rough energy
landscapes.
On the numerical side \cite{practical_guide2015}
so far, the models, which show such complex
behavior for ensembles with uncorrelated disorder, can
be treated only with an exponentially growing running time, let it
be Monte Carlo simulations \cite{newman1999} or
ground-state calculations \cite{opt-phys2001}.
This prohibits a sophisticated analysis. On the other
hand, for models where fast algorithms exist, e.g., random-field Ising systems
\cite{ogielsky1986},
two-dimensional spin glasses \cite{barahona1982b}, or matching problems
\cite{cormen2001}, the behavior
of uncorrelated or long-range power-law correlated disorder ensembles is
simple \cite{mezard1987matching,middleton2002,droplets2003,droplets_long2004,RFIM_correl2011,corr_2d_sg2021},
similar to a ferromagnet.

It is the purpose of the present paper to show that by using
more sophisticated disorder ensembles, in particular with suitable
correlations, indeed a complex behavior might be observed also for
models where fast and exact algorithms exist, allowing one to treat
very large system sizes. To be more precise,
here the directed polymer in a random medium (DPRM)
\cite{kardar1987,fisher1991,halpin-healy1995}
on a two-dimensional disordered
lattice was  studied. This model allows
for exact equilibrium
sampling of configurations for huge lattices with
even $N=10^9$ sites.
It is already known that directed polymers on random
trees exhibit one-step RSB \cite{derrida1988,mezard1991,derrida2016},
but on finite-dimensional lattices with ensembles of uncorrelated disorder,
no sign of complex behavior was found \cite{ueda2019}.
On the other hand, there are indications 
that by using another ensemble, originating from the Burger equation
\cite{ueda2015}, a complex behavior may be found, and also more general
approaches to complexity exist
\cite{franchini2019}. Motivated by these results,
in this work, lines or segments \cite{weinrib1983,meier2013}
of distinct disorder values  will be employed
in a novel way to the DPRM problem.
Here, a low temperature phase with a
broad distribution of overlaps and a ultrametric organization of the phase
space is present.

For the general two-dimensional case,
each realization of the model \cite{kardar1987,fisher1991}
is given by a  lattice
with $N=(L+1)\times (L+1)$ sites, open boundary conditions and
local quenched energy potential values
$\{V(x,y)\}$ for $x,y\in \{0,1,\ldots,L\}$. Directed polymers run from
$(0,0)$ to $(L,L)$ and contain $2L+1$ lattices sites
$P=\{(x_{\tau},y_{\tau})|\tau=0,\ldots, 2L\}$
and are located on adjacent lattice sites always moving towards
the final point $(L,L)$. Hence, 
for each ``time'' $\tau=x+y$, exactly one site is present in $P$,
and for  $(x,y)\in P$ with $x+y< 2L$ either $(x+1,y)\in P $ or $(x,y+1)\in P$. The energy of such a configuration is given
by the sum $E(P)=\sum_{(x,y)\in P} V(x,y)$ of the potentials of the
visited sites. The system is considered to be coupled
to a heat bath at temperature $T$, such that each valid polymer
exhibits a probability $e^{-E(P)/T}/Z$ with partition function
$Z=\sum_{P} e^{-E(P)/T}$. The model allows for each disorder realization
for a dynamic-programming,
or transfer-matrix,
calculation \cite{huse1985impurities,kardar1985,kardar1987}
of the partition function via
site-dependent partition functions with $Z(0,0)=e^{-V(0,0)/T}$ and
for $x,y=1,\ldots,L$: $Z(x,0)=Z(x-1,0)e^{-V(x,0)/T}$,
$Z(0,y)=Z(0,y-1)e^{-V(0,y)/T}$, and
$Z(x,y)=(Z(x-1,y)+Z(x,y-1))e^{-V(x,y)/T}$.
Note that $Z=Z(L,L)$. $Z$ can be calculated in time $O(L^2)$.
Furthermore, it is possible to sample polymer
configurations in exact
equilibrium by always starting with $P=P_0\equiv \{(L,L)\}$. Then one
adds further sites towards smaller times $\tau \to \tau-1$ as
follows: if the most recently added site is $(x,y)$, as next site
either $(x-1,y)$ is added to $P$,
with probability $Z(x-1,y)e^{-V(x,y)/T}/Z(x,y)$,
else site $(x,y-1)$, thus with probability $Z(x,y-1)e^{-V(x,y)/T}/Z(x,y)$.
If only one of the two sites is accessible, on the border of the lattice,
this single site is included in $P$. This process finishes when the
origin $(0,0)$ is reached. Each sampling requires only $O(L)$ steps.

\begin{figure}
  \includegraphics[width=0.3\linewidth]{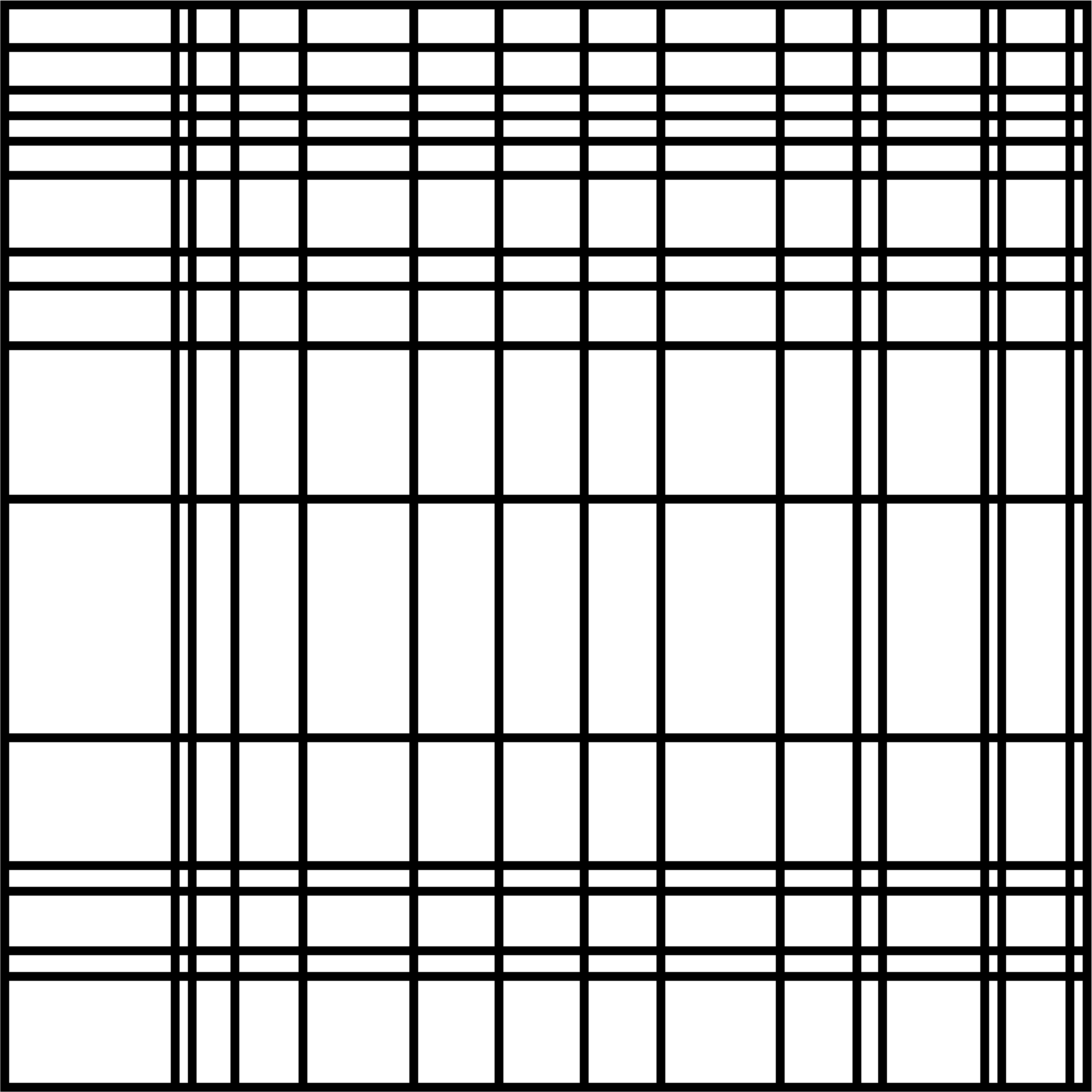}
  \includegraphics[width=0.3\linewidth]{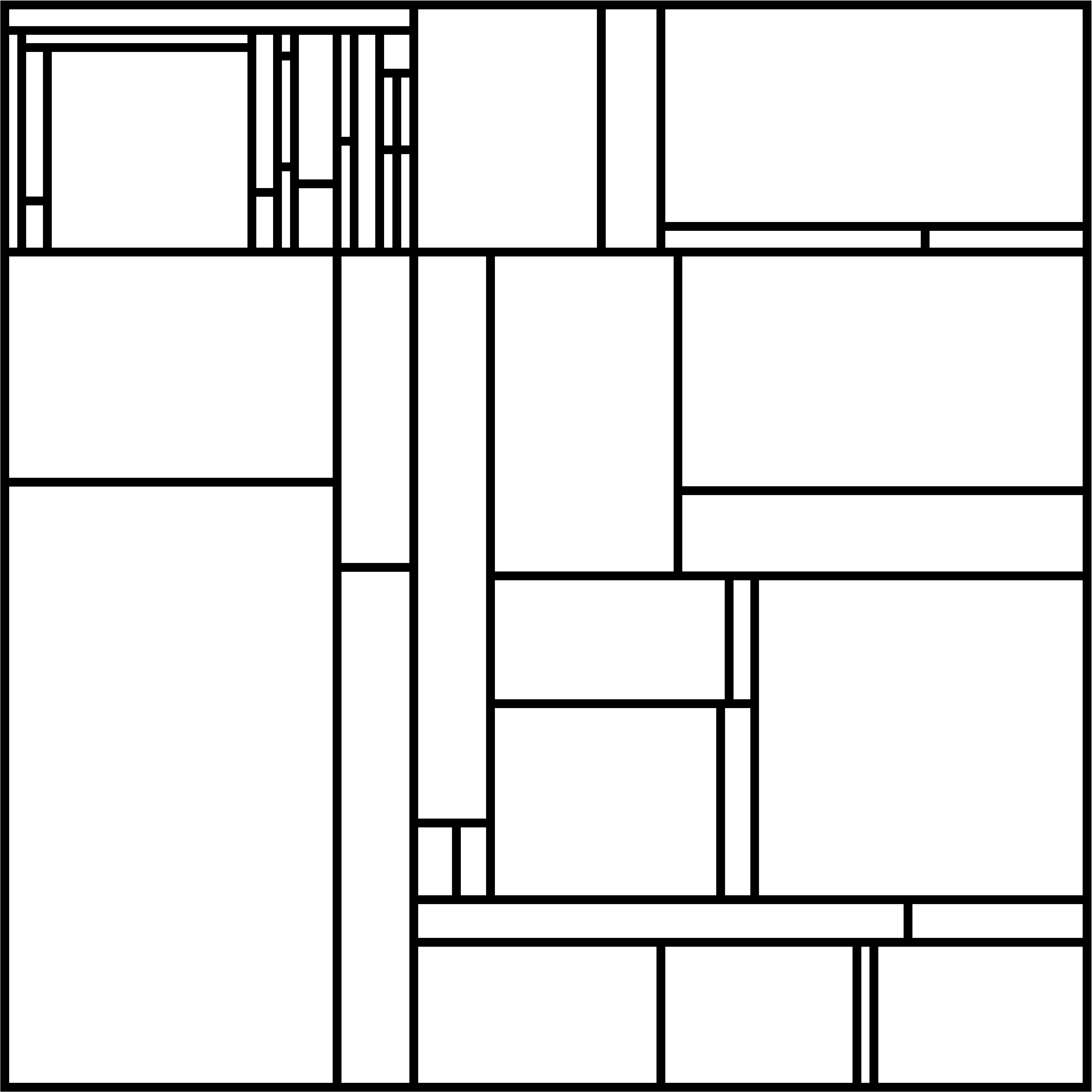}
  \includegraphics[width=0.3\linewidth]{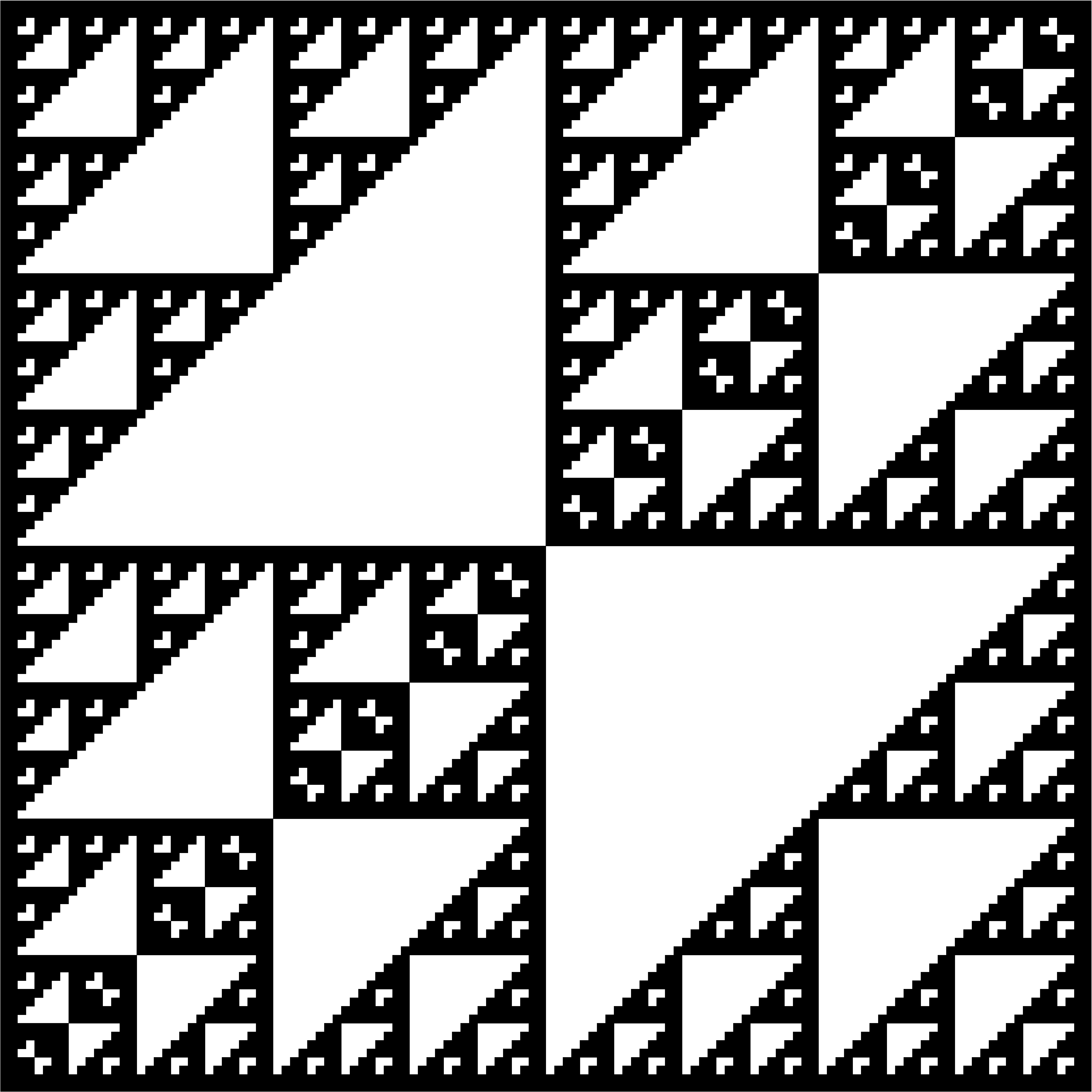}
  \caption{\label{fig:lattices} Examples
    of disorder realizations ($L=128$). White spaces correspond
    to potential $V=0$, black blocks to $V=-1$. Shown are (left)
    \emph{Hash} with 50 lines; (middle)
    \emph{Mondrian} with 50 lines; (right) \emph{Sierpinski} triangles. }
\end{figure}

Here two ensembles are considered where
most lattice sites have $V\equiv 0$ but in
addition segments or lines \cite{weinrib1983,meier2013}
on the lattice are introduced along which the potential has the same value
$V=-1$, favoring pinning of the polymer at low temperatures
\cite{giacomin2006}.
Third, an ``ensemble''
containing a single fractal structure of  potential values $-1$ and $0$
is investigated.

Here lattice sizes with $L=L_k=2^k$ are considered.
Each lattice exhibits
at the border a potential $V(x,y)\equiv -1$, i.e., for $x=0$, $x=L$, $y=0$
or $y=L$. There are more non-zero energy values,
which are chosen for three ensembles, see Fig.~\ref{fig:lattices}.
The ensembles \emph{Hash} \cite{weinrib1983,meier2013},
 \emph{Mondrian}, which is introduced in this work,
 and \emph{Sierpinski}, are defined as follows
\begin{itemize}
\item \emph{Hash}: A number $s$ of randomly chosen
  straight segments of length $L$ are added where $V\equiv -1$.
  This means, $l$ times a random point $(x_0,0)$ or $(0,y_0)$
  is selected and $V(x_0,y)\equiv -1$ or $V(x,y_0)\equiv -1$ is assigned for
  all $x,y\in\{1,\ldots,L-1\}$. 
  
\item \emph{Mondrian}: A set $D$ of straight segments is maintained,
  which contains initially the two segments $(0,0)\to(0,L)$
  and $(0,0)\to (L,0)$. Then $s$ times a segment is drawn with uniform
  probability $1/|D|$ from the current set $D$, without removing it.
  A site $(x_0,y_0)$
  is selected uniformly on this segment. Then a new segment is added to $D$
  which starts at the site $(x_0,y_0)$
  and runs, perpendicular to the selected segment, until any other segment
  from $D$ is hit. Finally, 
  all sites belonging to the segments in $D$
  obtain $V\equiv -1$ .
  
\item \emph{Sierpinski}: The discretized fractal Sierpinski structure with,
  for lattice size $L_k$, 
  $k-2$ recursion levels is embedded on the lattice. All sites belonging
  to Sierpinski triangles obtain $V\equiv -1$.
  \end{itemize}

For all ensembles, all other sites not having $V\equiv-1$,
  obtain $V\equiv 0$.
Here, for lattice size $L=L_k$,  $s_k=10(k-5)$
segments are inserted, respectively. Thus, the minimum meaningful lattice
size is $L_6=64$ for this study. Note that in Fig.~\ref{fig:lattices}
where $L=128=2^7$ instead of $s_{7}=20$ a higher number
of $s=50$ segments is used, for better visibility.

\begin{figure}
  \begin{minipage}[b]{0.32\linewidth}
  \includegraphics[width=0.99\linewidth,angle=180]
                  {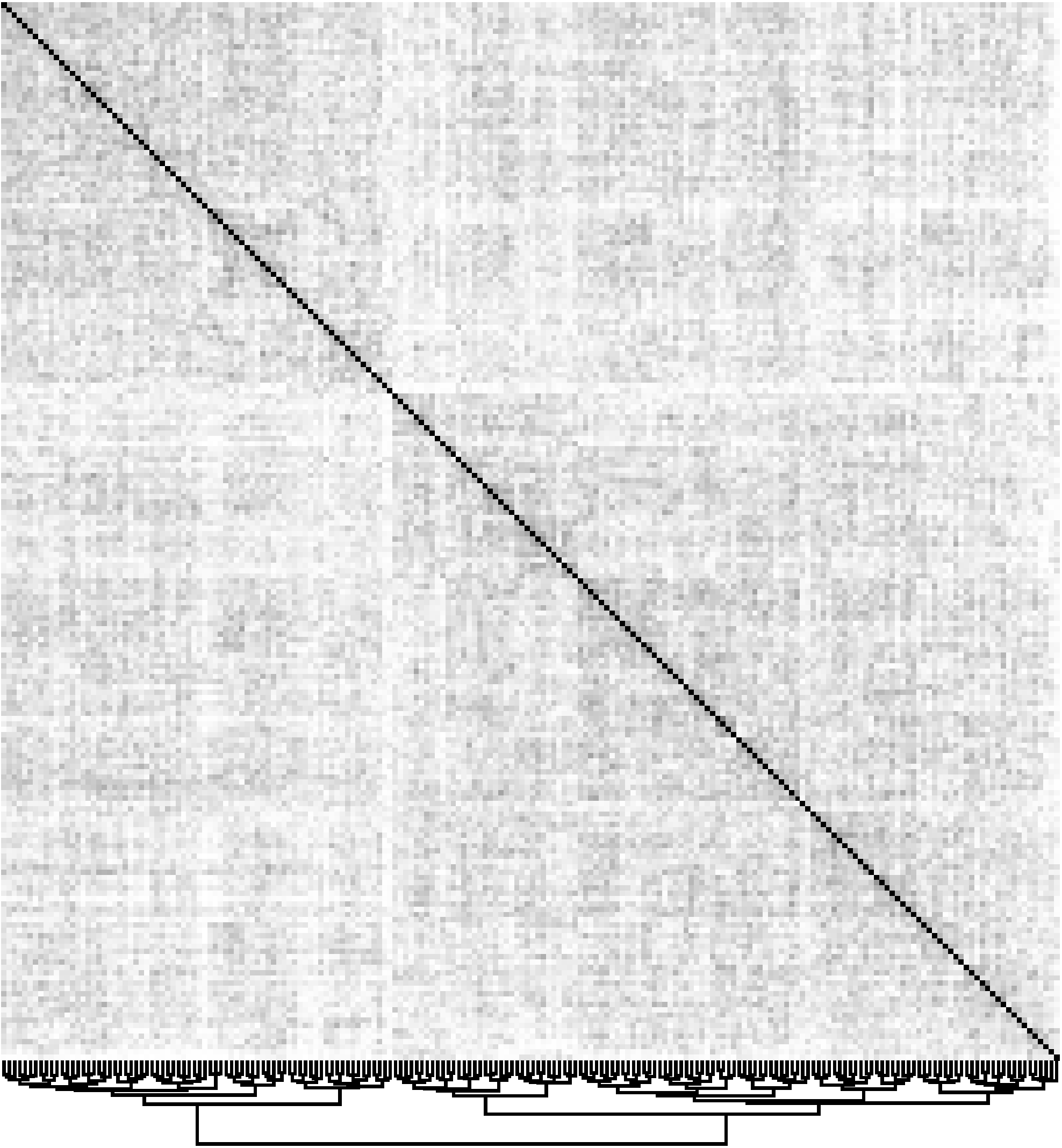}
\end{minipage}
\begin{minipage}[b]{0.32\linewidth}
    \includegraphics[angle=180,width=0.99\linewidth]
                  {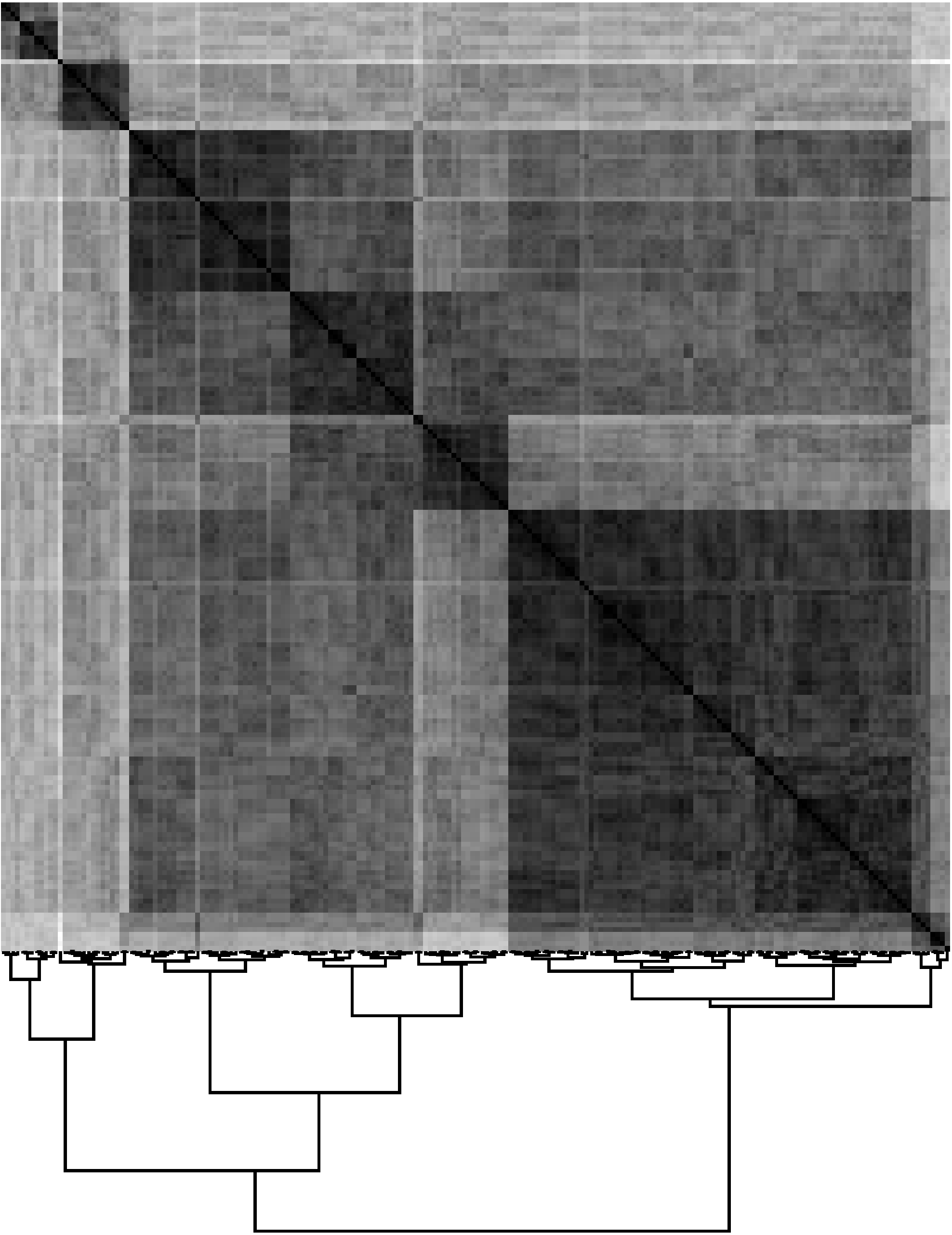}
\end{minipage}
\begin{minipage}[b]{0.32\linewidth}
  \includegraphics[angle=180,width=0.99\linewidth]
                  {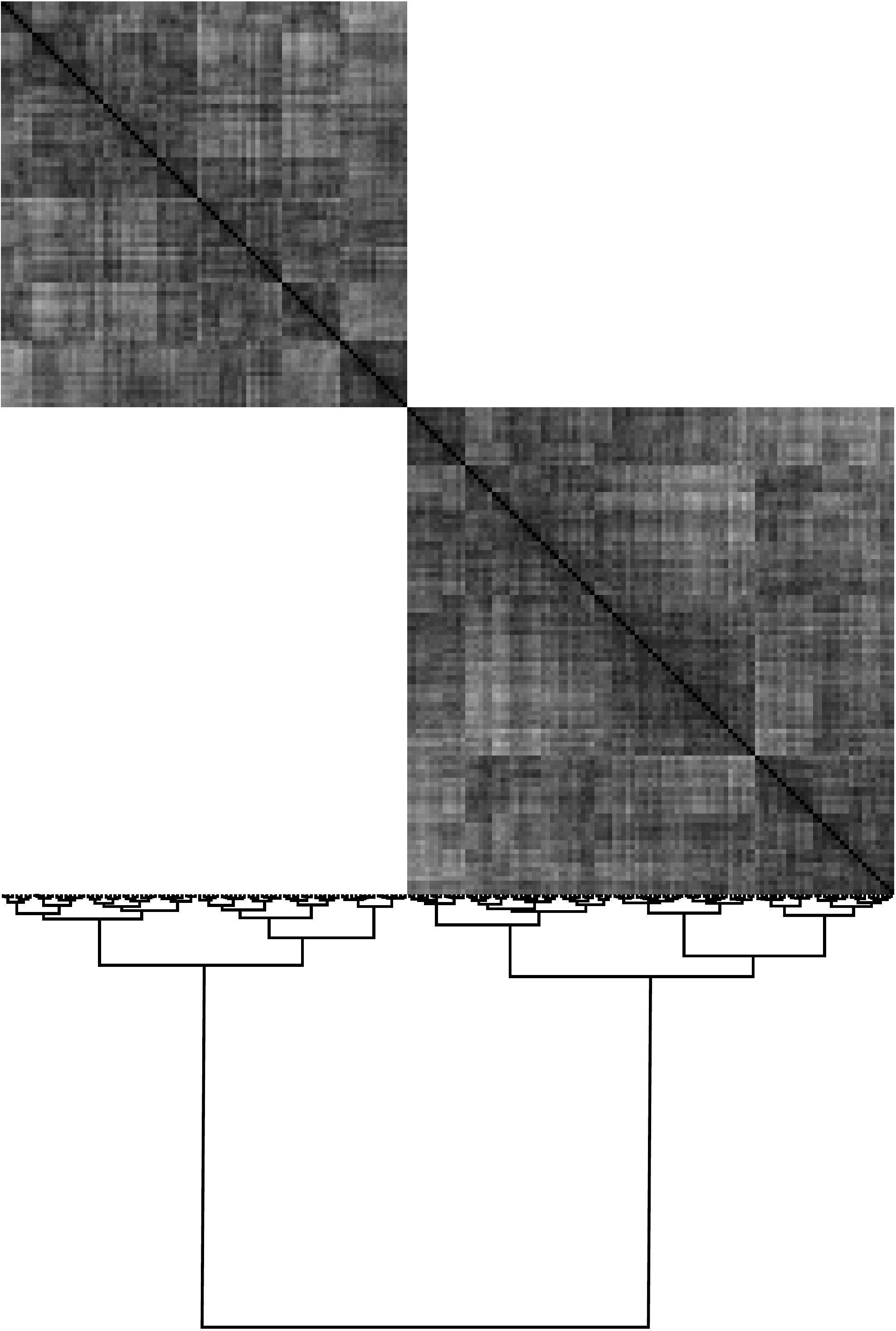}
\end{minipage}
  \caption{\label{fig:denograms} Examples
    for clustered overlap matrices with dendrograms showing
    the structures of the configurations space, each time 200 sampled
    for one realization ($L=16384, T=0.5$). A black dot means $q=1$
    while white corresponds to $q=0$.
     Shown are (left)
    \emph{Hash} with 90 lines; (middle)
    \emph{Mondrian} with 90 lines; (right) \emph{Sierpinski} triangles. }
\end{figure}

Each polymer configuration $P$ is characterized, first, by its energy $E(P)$
as defined above. This allows one to measure in equilibrium the
average energy $\langle E \rangle$ and the specific heat
$C(T)=(\langle E^2 \rangle -\langle E \rangle^2)/(NT^2)$,
for which one can also set up corresponding transfer-matrix equations
\cite{derrida1990}.
For the random-disorder ensembles a linear average of all
quantities over different realizations is performed,
not indicated by separate brackets here. To characterize
the model with respect to its energy landscape, the overlap $q$
between two polymers $P_1,P_2$ is used \cite{mukherji1994}, which is the
fraction of joint sites, i.e., $q_{12}\equiv
|P_1\cap P_2|/(2L+1)\in [0,1]$. By sampling
many polymers in equilibrium, evaluating all (or many) overlaps,
an approximation of the distribution $P(q)$ of overlaps is obtained.

To analyze the configuration space of these three ensembles,
different disorder realization were studied first at temperature $T=0.5$.
System sizes ranging from $L=64$ to $L=32768$ were considered.
A number of independent disorder realizations
ranging from 2000 for the smallest size to 500 for the largest size
were investigated.
For each disorder configuration $M=200$
independent polymer configurations were sampled in exact equilibrium.

The configuration space structure was analyzed by applying the
an \emph{agglomerative  clustering approach} of Ward
\footnote{
The clustering approach  \cite{ward1963,jain1988} operates on a set
of $M$ sampled configurations by initializing a set of $M$ clusters
each containing one configuration. One maintains pairwise distances
between all clusters, which are initially the distances between
the configurations. Then iteratively two clusters exhibiting the
currently shortest distance between them are selected and merged
to one single cluster, thereby reducing the cluster number by one.
For this new merged cluster, an updated distance to all other still
existing clusters have to be obtained. Here the update is done
with the approach of Ward \cite{ward1963}, which has
been used previously for the analysis of
disordered systems \cite{vccluster2004,1dchain_ultra2009,sat_cluster2010},
for more details see there.
The merging process is iterated until only one cluster is left.}.
The hierarchical structure obtained by the clustering
can be visualized by a tree, usually called
\emph{dendrogram}, where each branching corresponds to a subspace
of confirgurations,
see Fig.~\ref{fig:denograms}.
The sequence of configurations as located in the leafs
defines a partial order.
This order can be used to display the matrix of the pair-wise
overlaps  where the order of the rows and columns
is exactly given by the leaf order, see also Fig.~\ref{fig:denograms}.
For the \emph{Hash} ensemble, a rather gray uniform area is visible.
This indicates that the configuration space is rather uniform, like
a paramagnet. On the other hand, the matrices for the samples from
\emph{Mondrian} and \emph{Sierpinski} display a block-diagonal
structure, which is recursively visible inside the blocks as well.
This is an indication for a complex configuration space,
as it has been observed, e.g.,
for mean-field spin glass models \cite{1dchain_ultra2009}
or solution-space landscapes of optimization problems
\cite{vccluster2004, sat_cluster2010}.

\begin{figure}
  \includegraphics[width=0.48\linewidth]{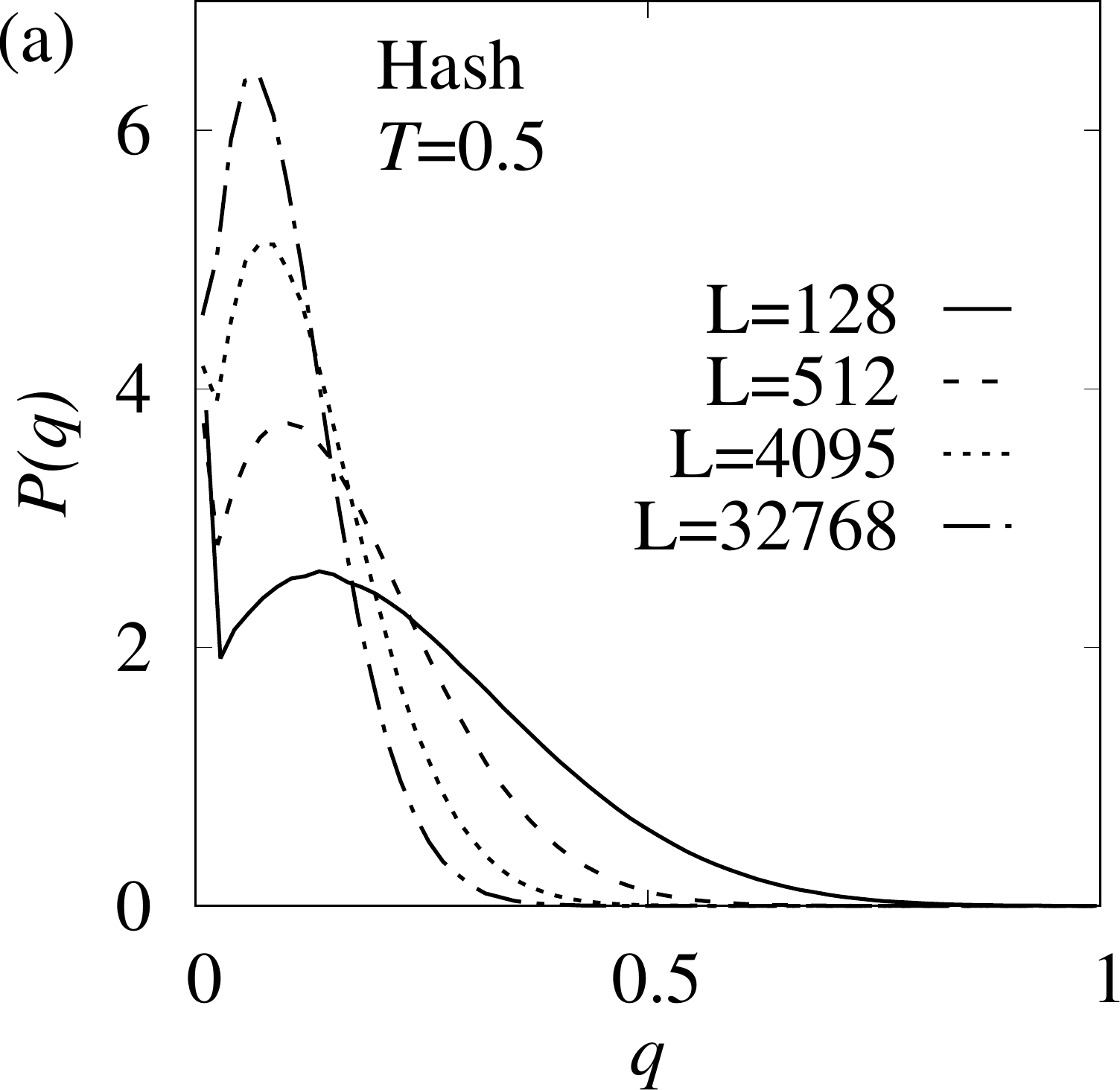}
  \includegraphics[width=0.48\linewidth]{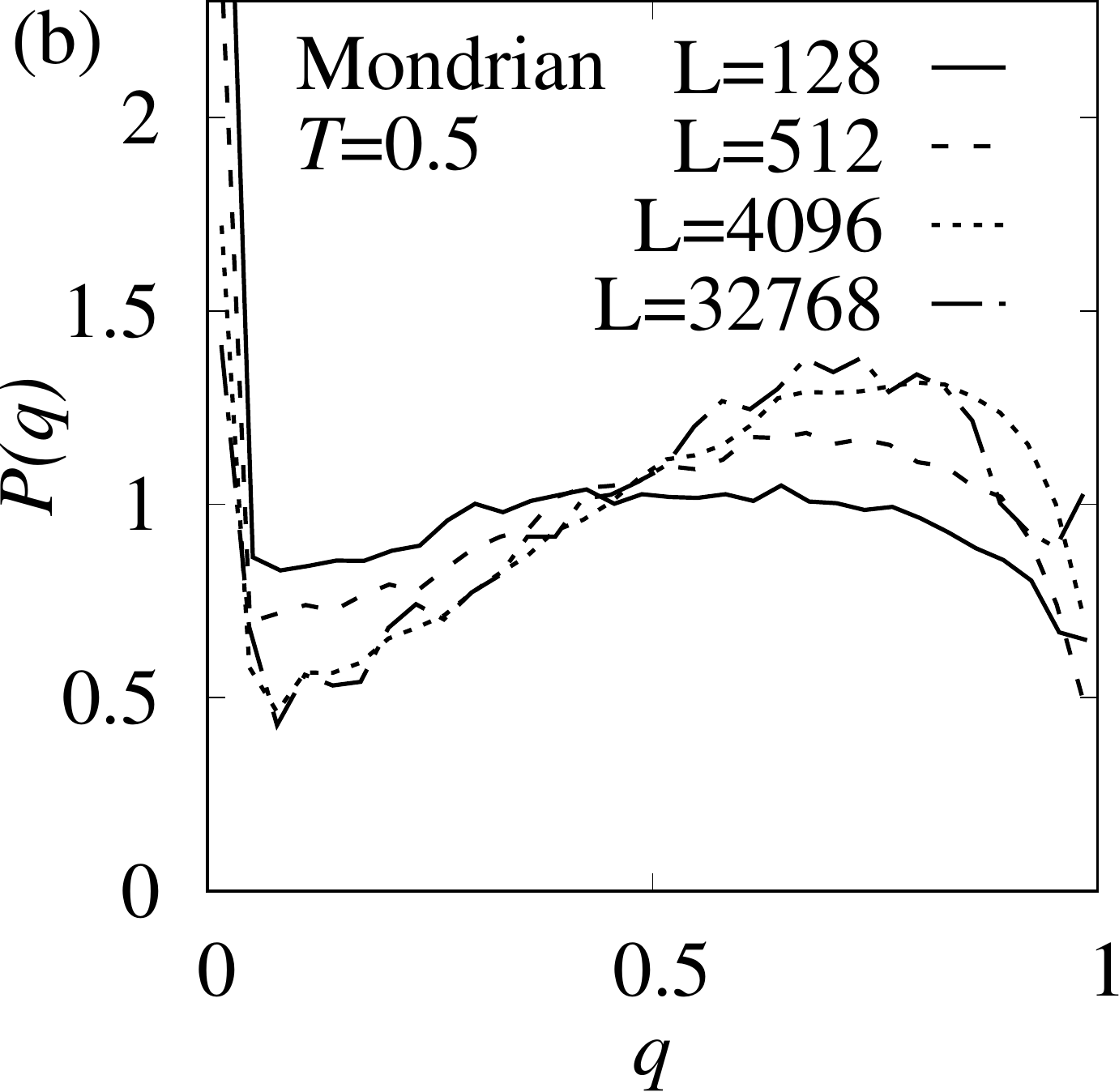}

  \includegraphics[width=0.48\linewidth]{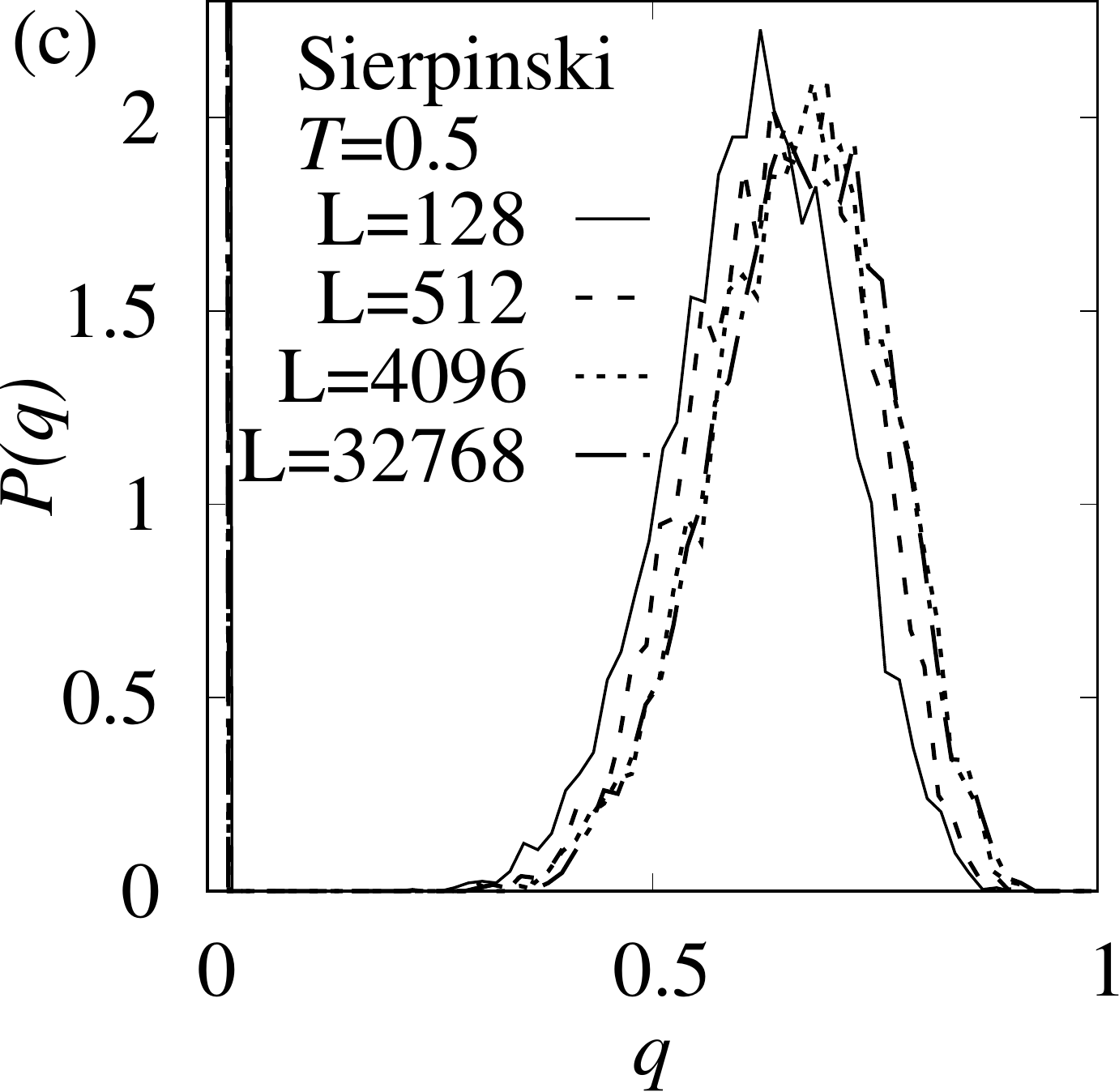}
  \includegraphics[width=0.48\linewidth]{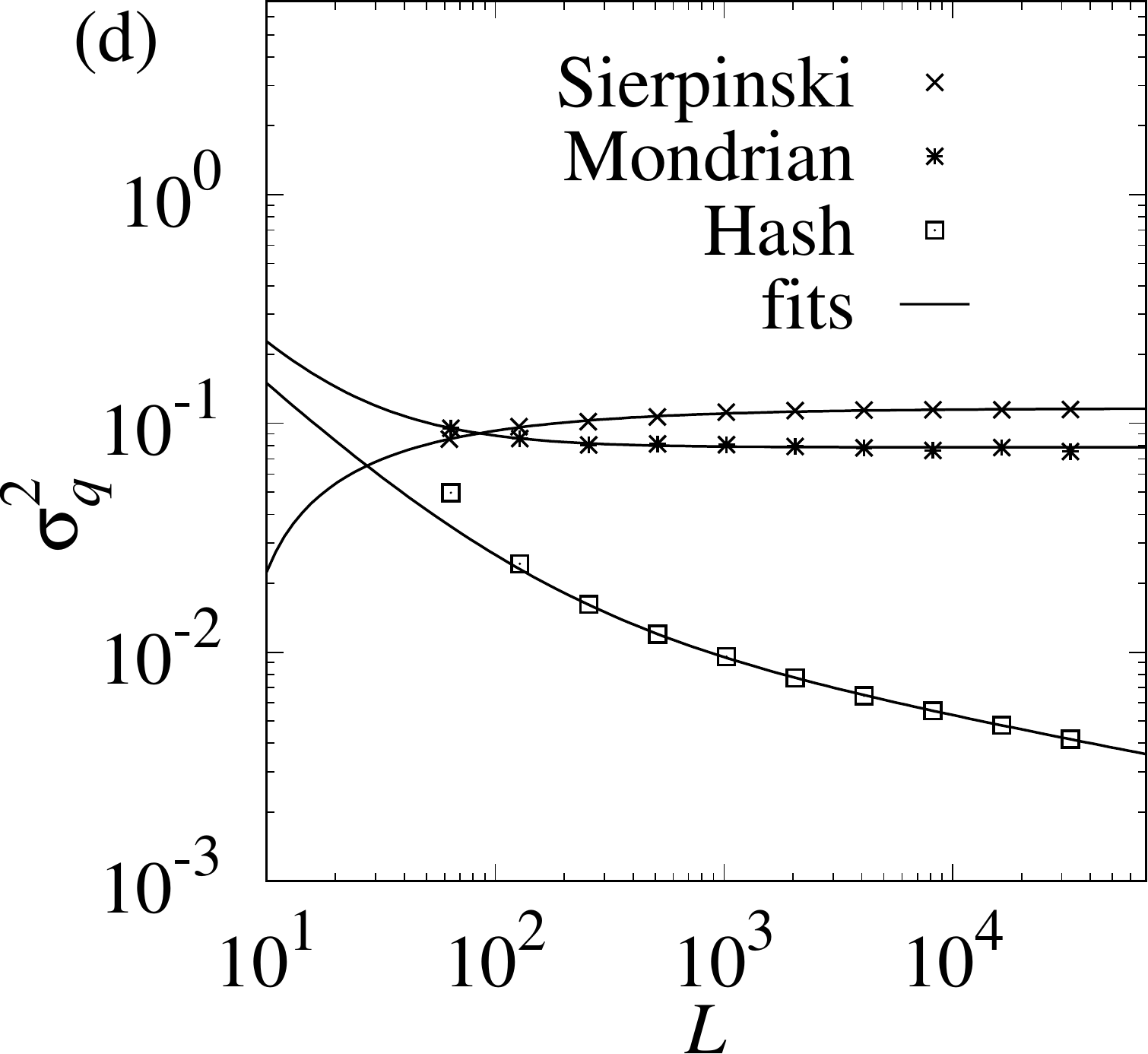}
  \caption{\label{fig:Pq} Distribution $P(q)$ of overlaps at temperature
    $T=0.5$ for four different system sizes. The results are for (a)
    \emph{Hash},  (b)
    \emph{Mondrian}, and  (c) \emph{Sierpinski} triangles.
    In (d) the variance $\sigma^2_q$ of
    these distributions for the three ensembles is shown as function
    of the system size $N$,
  together with fits (see text).}
\end{figure}

In Fig.~\ref{fig:Pq} the distributions of overlaps are shown.
For the \emph{Hash} case, the distribution gets strongly
narrower, indicating a convergence
to $P(q)=\delta(q-q_0)$, which corresponds to a trivial configuration landscape.
From a fit of the mean as function of $L$ to a power-law plus constant
$q_0$, a value of $q_0=0.081(3)$ was obtained.
On the other
hand, for the other two ensembles $P(q)$ seems to converge to a broad
distribution for $q>0$ plus a delta-peak at $q=0$ with some weight $w_0$,
which
accounts for polymers having different paths right from the start.
For for the \emph{Sierpinski} ensemble the data exhibits a convergence
to $w_0=0.5$.
This is compatible with the structure of the lattice, since 
at the starting site the paths either go down or right and never meet again,
thus half of the pairs have zero overlap.
For the \emph{Mondrian} ensemble, a much smaller limiting
zero-overlap peak-weight $w_0\approx 0.04$ is found, i.e., most
of the overlap distribution is located in the non-trivial part.
Also shown in Fig.~\ref{fig:Pq} are the variances $\sigma_q^2$ of the
distributions of overlaps for the three ensembles. For the \emph{Mondrian}
and the \emph{Sierpinski} ensembles, the variance seems to converge
to finite values in the $L\to\infty$ limit. This is confirmed by good fits
for $L>100$ of the data to functions of the form $\sigma(N)=\sigma_{\infty}+aL^{-b}$
which lead to clear non-zero values $\sigma_{\infty}=0.1163(8)$ for the
\emph{Sierpinski} ensemble and $\sigma_{\infty}=0.0785(7)$
for the Mondrian ensemble.
Thus, for these two ensembles the distribution of overlaps remains
broad at low temperature in the thermodynamic limit $L\to\infty$ indicating
a complex phase space structure.
The variance for the \emph{Hash} ensemble exhibits a positive curvature
in the log-log plot, which could also be taken as indication for
a complex structure. Nevertheless, here each
polymer path can
be decomposed in many sub paths with a high degree of independence,
which speaks in favor of a simple configuration-space structure. Indeed,
a limiting zero width is compatible: When fitting for $L>100$ a
power-law with a correction term,
$\tilde \sigma(N)=aL^{-b}(1+eL^{-d})$, a good fit is obtained as well,
as shown in the figure.

\begin{figure}
  \includegraphics[width=0.48\linewidth]{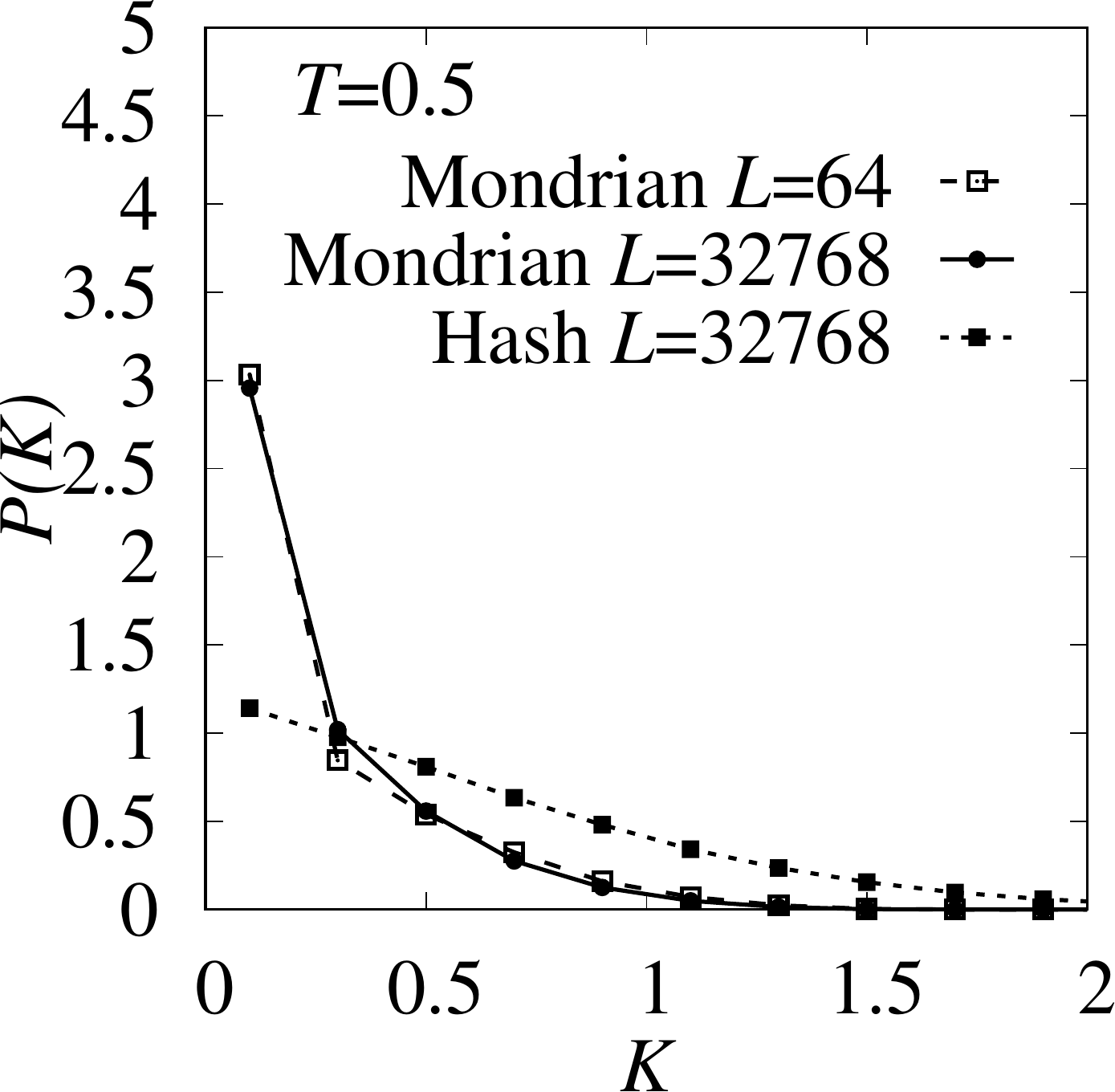}
  \includegraphics[width=0.48\linewidth]{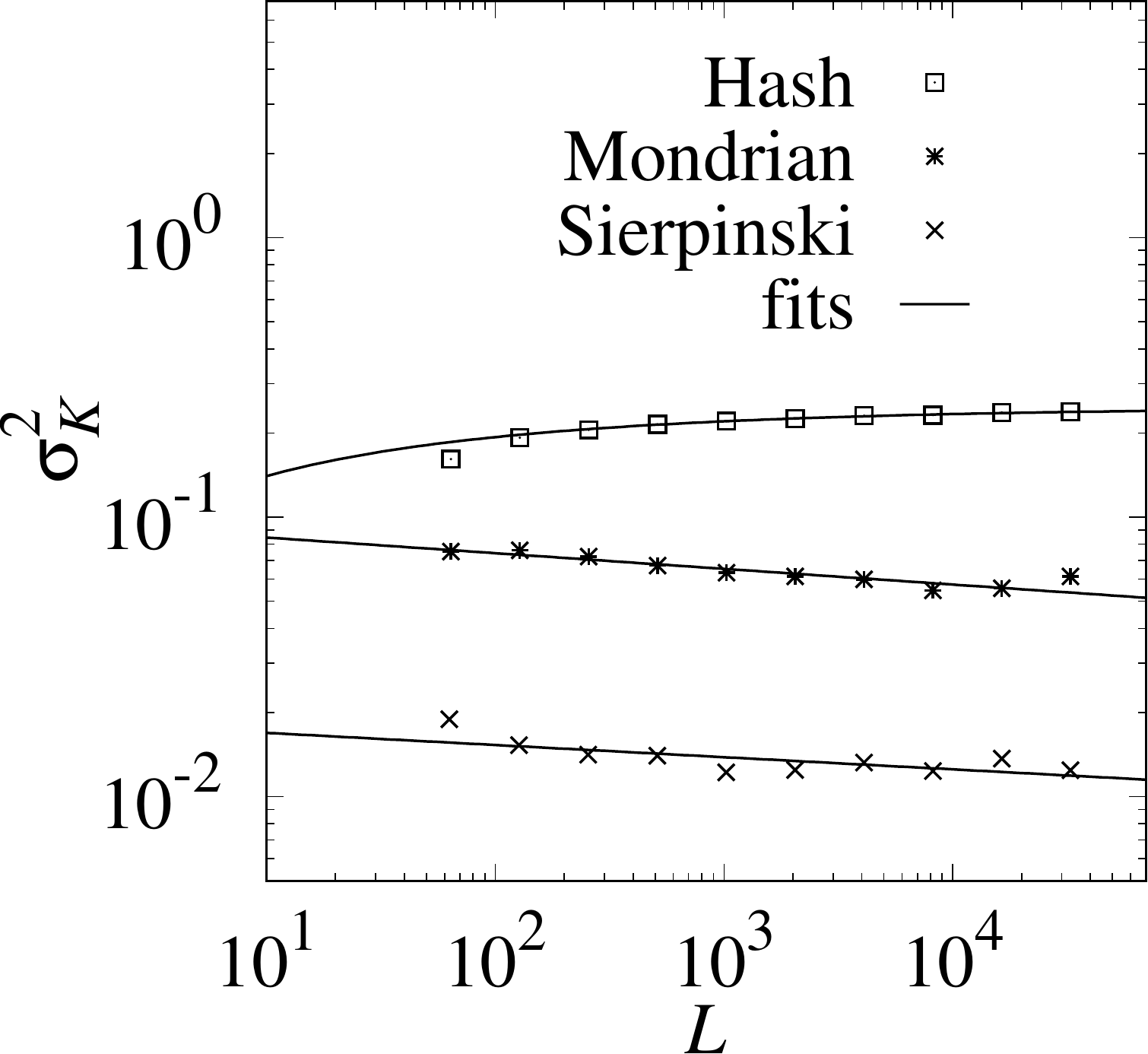}
  \caption{\label{fig:PK} (left) Sample distributions $P(K)$
    of the ultrametricity measure for the
    \emph{Mondrian}, and \emph{Hash} ensembles.
    (Right) variance $\sigma^2_K$ of
    these distributions for the three ensembles as function
    of the system size $N$,
  together with fits (see text).}
\end{figure}

A hierarchical configuration space, like for the SK model, is characterized
by an \emph{ultrametric} structure \cite{rammal1986}, i.e.,
an underlying tree. To
characterize ultrametricity, one considers triples of configurations
$P_1,P_2,$ and $P_3$ and their mutual overlaps $q_{12}, q_{13},$
and $q_{23}$ which are, without loss of generality, ordered
such that $q_{12}\le q_{13} \le q_{23}$. For a true ultrametric
space, for an infinite system size, $q_{12}=q_{13}$ would hold.
To characterize the emergence of ultrametricity here, the quantity
$K=(q_{13}-q_{12})/\sigma_q$ is used \cite{1dchain_ultra2009},
where $\sigma_q$ is the width
of the overlap distribution $P(q)$. For a non-trivial ultrametric
organization, the distribution $P(K)$ should converge to a delta-function
$\delta(K)$, i.e., a variance $\sigma^2_K$ which converges to zero.
In the left of Fig.~\ref{fig:PK} samples for $P(K)$
 are shown for \emph{Mondrian} and \emph{Hash}
ensembles. The former one exhibits a slight change towards smaller values
of $K$ when increasing the system size $L$. For the latter one,
the distribution is much broader, also for the largest considered size.
This is confirmed by the behavior of the
variance $\sigma_K^2$ of these distributions as function of the system size.
The data is compatible with a gentle power-law decreases, shown as
straight lines, for the
\emph{Mondrian} and the \emph{Sierpinski} ensembles. This can be
expected for the fractal \emph{Sierpinski} ensemble since it has an obvious
hierarchical structure. Note that the convergence even in this obvious
ultrametric case is slow, as it was also observed for long-range
spin glasses exhibiting RSB \cite{1dchain_ultra2009}.
Thus, the data indicates that also the \emph{Mondrian}
ensemble exhibits ultrametricity as well. Also, the variance
seems to converge to a constant for the \emph{Hash} ensemble,
compatible with the absence ultrametricity, and expected
because of the simpler distribution of overlaps.

\begin{figure}
  \includegraphics[width=0.48\linewidth]{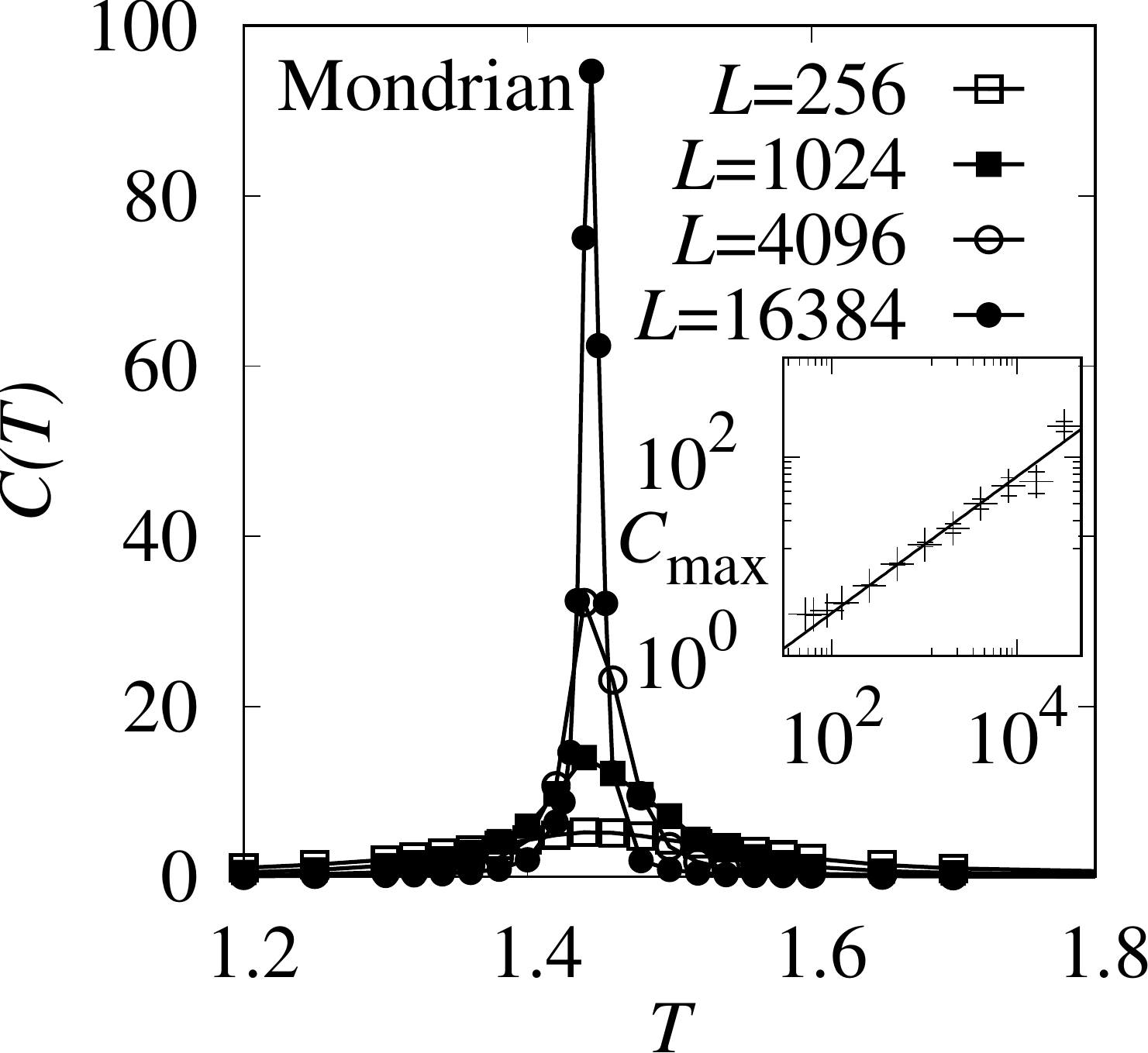}
  \includegraphics[width=0.50\linewidth]{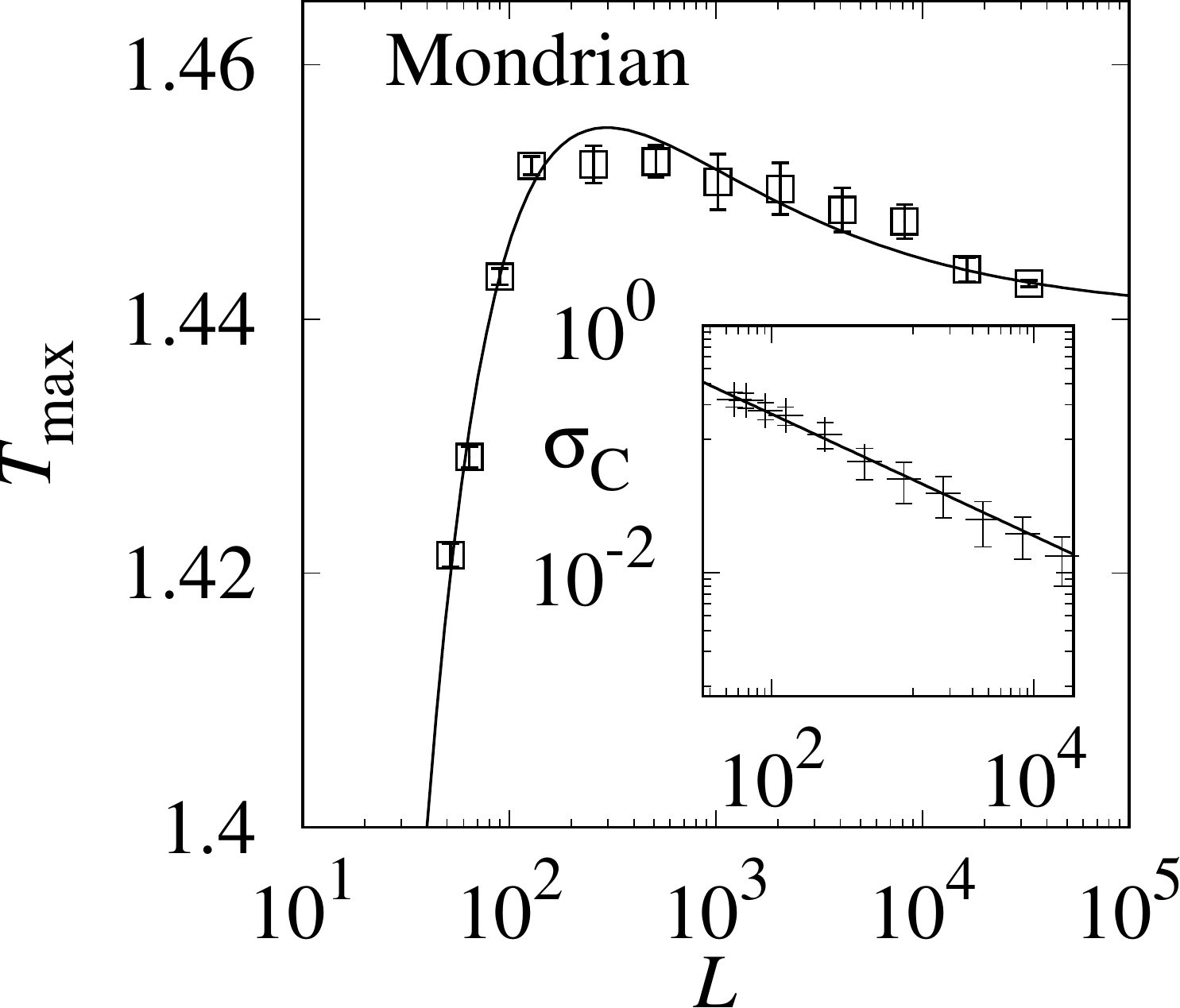}
  \caption{\label{fig:C} (left) Specific heat $C(T)$ as function of
    temperature for the
    \emph{Mondrian} ensemble and four different system sizes. The
    inset shows the peak height $C_{\max}$ as function of the system size.
  (right) Position $T_{\max}$ of the peak of $C(T)$ as function of
    system size $L$ and in the inset the width $\sigma_{C}$ of the
    peak. The lines for $C(T)$ are guide to the eyes, the other lines
  display fits (see text).}
\end{figure}

In order to study the temperature dependence \cite{derrida1990}
of an  ensemble with complex behavior, for \emph{Mondrian},
a large number of simulations was
performed. Note that similar simulations for the Sierpinski model exhibited
hard to analyze discotinuities and are thus not presented here.
Lattice sizes $L\le 16384$ for many temperatures $T\in [0.1,3]$,
plus for $L=32768$ for few temperatures near the estimated critical
point were considered with  
the number of disorder samples between 500 and 1000. 
In the left of Fig.~\ref{fig:C}  examples for the specific heat $C(T)$ 
behavior is shown. Clearly 
peaks are visible near $T\approx 1.4$, growing and narrowing
 with increasing system size, indicating a phase transition.
For a second order phase transition \cite{stanley1971,goldenfeld1992,cardy1996,yeomans2002}
one would expect that the specific heat scales as
\begin{equation}
  \label{eq:c}
  C(T,L)=L^{\alpha/\nu} \tilde c( (T-T_c)L^{1/\nu})\,,
  \end{equation}
with a size-independent function $\tilde c()$
and critical exponents $\nu$,
describing the divergence of the correlation length, and $\alpha$ describing
the divergence of the specific heat. Indeed,
the height of the peak 
follows clearly a power law $C_{\max}(L)\sim L^{\alpha/\nu}$, see inset
of the left Fig.~\ref{fig:C}. A fit to this power law
results in $\alpha/\nu= 0.69(2)$. 

The position $T_{\max}$
of the peak was estimated
by fitting Gaussians near the peak. The position as a function of the
system size is shown in the right of Fig.~\ref{fig:C}. Only a weak,
third-digit significant, but non-monotonous size dependence is visible.
Equation (\ref{eq:c}) means that
scaling of the peak position leads to a
leading behavior $T_{\max}(N)-T_c\sim L^{-1/\nu}$.  Nevertheless, fitting just
a power law does not work well, even when restricting to larger sizes.
 On the other hand, Eq. (\ref{eq:c}) also
concerns the shape of the specific heat, i.e., the width of the peak
region should also scale like $L^{-1/\nu}$. The width, as obtained also
from the Gaussian fits, shows indeed a clear power law. A fit to a
power law,
yielded $\nu=-2.02(8)$.
When fixing $\nu$ to this value, a fit to a power-law with correction
$T_{\max}(N)=T_c+cL^{-1/\nu}(1+dL^{-\omega})$ yields a reasonable fit,
see Fig.~\ref{fig:C}, with $T_c=1.439(8)$. With this value of
$\nu$, a rather large value of $\alpha\approx 1.4$ results,
which could indicate that actually a first-order phase transition is
behind the seen data \footnote{In renormalization group studies on obtains
  for a $d$-dimensional system  $\alpha=2-d/y_1$. For a first-order phase
  transition $y_1=d$ holds   \cite{nienhuis1975}, which leads to $\alpha=1$
  which is large compared to typically observed values}. This
is compatible with the observed discontinities
of  the related Sierpinski lattice.


\begin{figure}
  \includegraphics[width=0.48\linewidth]{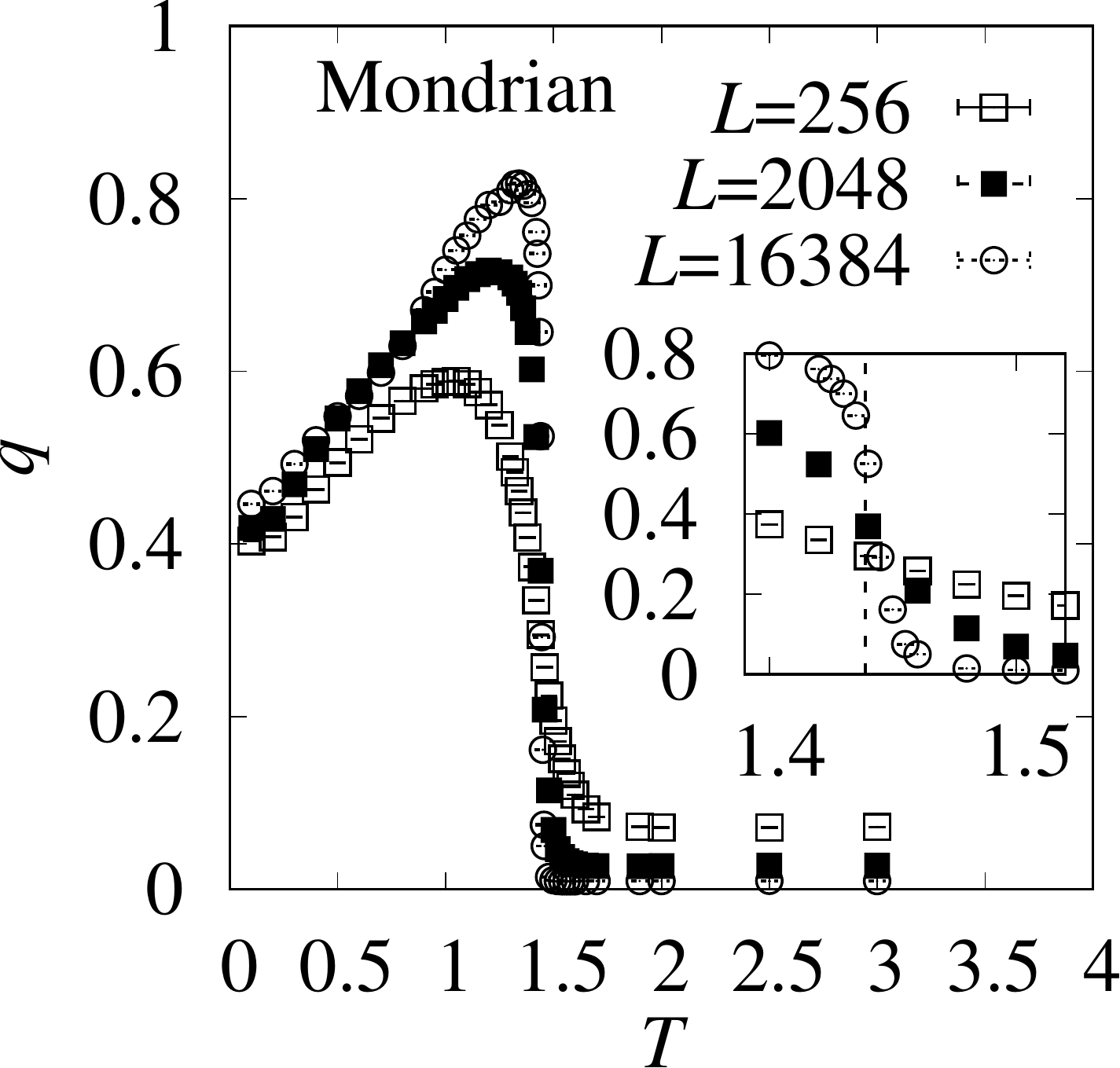}
  \includegraphics[width=0.48\linewidth]{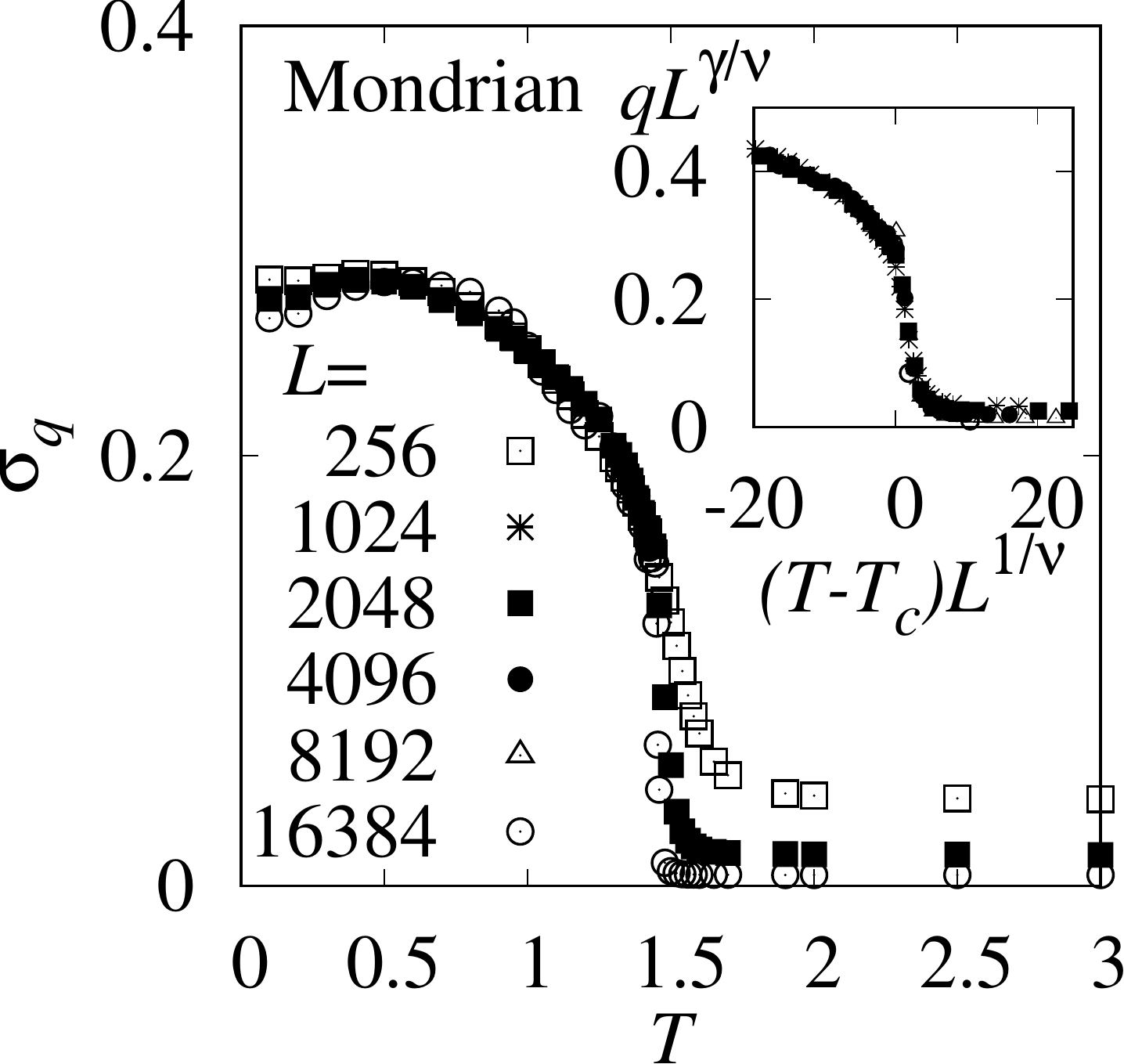}
  \caption{\label{fig:q} \emph{Mondrian} ensemble and different system
    sizes in the (left) the mean overlap $q(T)$ as function of
    temperature.
    The
    inset shows the data near the estimated transition point,
    indicated by a vertical dashed line. (right) mean width $\sigma_q(T)$
    of the overlap distribution, for three sample system sizes
    The inset shows the rescaled data for system sizes $L \ge 1024$.}
\end{figure}

The average overlap $q(T)$ is shown in the left of Fig.~\ref{fig:q}. At low
temperatures  $T>T_c$, the average overlap is non-zero. The curves for different system sizes cross near $T_c$ and
just below $T_c$ the average overlap grows with the system size. This is an
unusual behavior when comparing, e.g., with a ferromagnet. A data collapse
(not shown) leads to an unphysical negative critical exponent. Note that
also the average squared overlap (not shown) exhibits this behavior.
The average width $\sigma_q(T)$ of the overlap
distribution is shown in the right of Fig.~\ref{fig:q}.
The data can be rescaled reasonably well, see inset, according to
$\sigma_q(T,L)=L^{-\gamma/\nu} \tilde \sigma ((T-T_c)L^{1/\nu})$
when using the values $T_c=1.439$, $\nu=2.02$ obtained already
and estimating $\gamma/\nu=0.07(2)$.  The smallest system sizes are
excluded from the collapse due to too large finite-size corrections. 
$L=32768$ is not included here due to bad statistics.

To conclude, it was shown that some specific ensembles of the disorder
for random polymers on a two-dimensional lattice, at low temperatures
exhibit a complex
hierarchical organization of the phase space,
similar to RSB. In contrast to other models
exhibiting complex behavior, the present models allows for fast and exact
sampling at arbitrary temperatures, i.e., to study large system
in true equilibrium. This may open a path, by just using
suitably correlated disorder ensembles, to study in a numerically
convenient way complex behavior. This may be done for other disorder
ensembles, other lattice dimensions or even other models
where exact equilibrium sampling is possible.

\acknowledgments{The author thanks A. Peter Young,
  Hendrik Schawe and Phil Krabbe
  for critically reading  the manuscript and useful discussions.
  The simulations were performed at the 
  the HPC cluster CARL, located at the University of Oldenburg
  (Germany) and
    funded by the DFG through its Major Research Instrumentation Program
    (INST 184/157-1 FUGG) and the Ministry of
    Science and Culture (MWK) of the
    Lower Saxony State. 
}

\bibliography{alex_refs,polymer}

\end{document}